# Fibonacci-Like Polynomials Produced by m-ary Huffman Codes for Absolutely Ordered Sequences

**Alex Vinokur**
Holon, Israel
alexvn@barak-online.net
alex.vinokur@gmail.com

**Abstract**. A non-decreasing sequence of positive integer weights $P = \{p_1, p_2, \ldots, p_n\}$ ($n = N*(m-1) + 1$, $N$ is number of non-leaves of $m$-ary tree) is called *absolutely ordered* if the intermediate sequences of weights produced by $m$-ary Huffman algorithm for initial sequence $P$ on $i$-th step satisfy the following conditions $p_m^{(i)} < p_{m+1}^{(i)},\ i = \overline{0, N-2}$. Let $T$ be an $m$-ary tree of size $n$ and $M=M(T)$ be a set of such sequences of positive *integer* weights that $\forall P \in M$ the tree $T$ is the $m$-ary Huffman tree of $P$ ($|P|=n$). A sequence $P_{\min}$ of $n$ positive integer weights is called a *minimizing* sequence of the $m$-ary tree $T$ in the class $M$ ($P_{\min} \in M$) if $P_{\min}$ produces the minimal Huffman cost of the tree $T$ over all sequences from $M$, i.e., $E(T,P_{\min}) \le E(T,P)\ \forall P \in M$. **Theorem 1**. A *minimizing* absolutely ordered sequence of size $n = N*(m-1) + 1$ for the maximum height $m$-ary Huffman tree ($m > 1$) is

$$Pmin_{abs}(N, m) =$$
$$\{ G_0(m-1), \underbrace{G_1(m-1),\ldots,G_1(m-1)}_{(m-1)\text{ times}}, \underbrace{G_2(m-1),\ldots,G_2(m-1)}_{(m-1)\text{ times}},\ldots, \underbrace{G_N(m-1),\ldots,G_N(m-1)}_{(m-1)\text{ times}} \},$$

where $G_0(m) = 1$, $G_1(m) = 1$, $G_2(m) = 2$, $G_i(m) = G_{i-1}(m) + m*G_{i-2}(m)$ when $i = \overline{2, N}$ ■
Polynomials $G_i(x)$ are called Fibonacci-like polynomials. **Theorem 2**. The cost of maximum height $m$-ary Huffman tree $T$ of size $n = N*(m-1) + 1$ for the minimizing absolutely ordered sequence $Pmin_{abs}(N, m)$ is

$$E(T, Pmin_{abs}(N, m)) = \frac{G_{N+4}(m-1) - 2}{m-1} - (N+3).\ ■$$

Samples of Fibonacci-like polynomials and costs of maximum height $m$-ary Huffman trees are shown.

## 0. Preface

Absolutely ordered and $k$-ordered sequences for *binary* Huffman trees those have maximum height have been investigated in [1] and [2]. In this article the generalization of absolutely ordered sequences for $m$-ary Huffman trees those have maximum height is considered.

# 1. Main Conceptions and Terminology

## 1.1. m-ary trees

A (strictly) *m-ary tree* is an oriented ordered tree where each nonleaf node has exactly $m$ children (siblings). *Size of an m-ary tree* is the total number of *leaves* of this tree. Let $N$ be number of nonleaves (internal nodes), $n$ be number of leaves of $m$-ary tree. Number of leaves in $m$-ary tree satisfies the following condition

$$n = N*(m-1) + 1. \qquad (1)$$

An $m$-tree ($m \geq 2$) is called *elongated* if at least ($m$-1) of any $m$ sibling nodes are leaves. An elongated binary tree of size $n$ has maximum height among all binary trees of size $n$. An elongated $m$-ary tree is called *left-sided* if only the *left* node in each $m$-tuple of sibling nodes can be nonleaf.

A $m$-ary tree is called *labeled* if a certain positive integer (weight) is set in correspondence with each leaf.

**Definition**. Let $T$ be an $m$-ary tree with positive weights $P=\{p_1,.., p_n\}$ at its leaf nodes. The *weighted external path length* of $T$ is

$$E(T,P) = \sum_{i=1}^{n} l_i p_i$$

where $l_i$ is the length of the path from the root to leaf $i$.

## 1.2. Generalized *m*-ary Huffman algorithm

**Problem definition**. Given a sequence of $n$ positive weights $P=\{p_1,\ldots, p_n\}$, $(n-1) = 0$ (**mod**$(m-1)$). The problem is to find $m$-ary tree $T_{\min}$ with $n$ leaves labeled $p_1,..., p_n$ that has minimum weighted external path length over all possible $m$-ary trees of size $n$ with the same sequence of leaf weights. $T_{\min}$ is called the $m$-ary Huffman tree of the sequence $P$; $E(T,P_{\min})$ is called the Huffman cost of the tree $T$.

The problem was solved for binary trees by Huffman algorithm [3]. That algorithm can be generalized for $m$-ary trees. A generalized Huffman algorithm builds $T_{\min}$ in which each leaf (weight) of $m$-ary tree is associated with a (prefix free) codeword in alphabet $\{0, 1,\ldots, m-1\}$.

*Note*. A code is called a prefix (free) code if no codeword is a prefix of another one.

***m*-ary algorithm description** (in the reference to the discussed issue).

Algorithm input. A non-decreasing sequence of positive weights

$P =\{p_1, p_2,\ldots, p_n\}$ $p_k \leq p_{k+1}, k = \overline{1,n-1}$; $n = N*(m-1)+1$, where $N$ is number of non-leaves.

Algorithm output. The sum of all the weights.

The algorithm is performed in $N$ steps. $i$-th step ($i = \overline{1,N}$) is as follows.

- $i$-th step input. A non-decreasing sequence of weights of size $n-(m-1)*(i-1)$.

$P^{(i-1)}=\{ p_1^{(i-1)}, p_2^{(i-1)},..., p_{n-(m-1)*(i-1)}^{(i-1)} \}$ ( $p_k^{(i-1)} \leq p_{k+1}^{(i-1)}; k = \overline{1,n-(m-1)*(i-1)-1}$); $|P^{(i-1)}|=n-(m-1)*(i-1)$.

- $i$-th step method. Build a sequence

$$\{ p_1^{(i-1)} + p_2^{(i-1)} + ... + p_m^{(i-1)}, p_{m+1}^{(i-1)},..., p_{n-(m-1)*(i-1)}^{(i-1)} \}$$

  and sort its.

- $i$-th step output. A non-decreasing sequence of weights of size $n-(m-1)*i$.

$P^{(i)}=\{ p_1^{(i)}, p_2^{(i)},..., p_{n-(m-1)*i}^{(i)} \}$ ( $p_k^{(i)} \leq p_{k+1}^{(i)}; k = \overline{1,n-(m-1)*i-1}$); $|P^{(i)}|=n-(m-1)*i$.

*Note 1*. $P^{(0)}$ is an input of $m$-ary Huffman algorithm, i.e.,

$$p_k^{(0)} = p_k \ (k = \overline{1,n}).$$

*Note 2*. If an input sequence on *i*-th step(s) of the algorithm satisfies condition

$$p_m^{(i)} = p_{m+1}^{(i)} (0 \le i \le N-2),$$

then several *m*-ary Huffman trees can result from initial sequence *P* of weights, but the weighted external path length is the same in all these trees.

Let $P = \{p_1, p_2, p_3, \ldots, p_n\}$ be a sequence of size *n* for which the *m*-ary Huffman tree is *elongated*. Then according to generalized *m*-ary Huffman algorithm

$$p_1^{(i)} + p_2^{(i)} + \ldots + p_m^{(i)} \le p_{2m}^{(i)}, i = \overline{0, N-2}. \quad (2)$$

## 2. Main Results

### 2.1. Minimizing absolutely ordered sequence of the elongated *m*-ary Huffman tree

Let *T* be an *m*-ary tree ($m > 1$) of size *n* (i.e., $n = N*(m-1) + 1$, where *N* is number of non-leaves and *n* is number of leaves) and $M=M(T)$ be a set of such sequences of positive *integer* weights that $\forall P \in M$ the tree *T* is the *m*-ary Huffman tree of *P* ($|P|=n$).

**Definition**. A sequence $P_{\min}$ of *n* positive *integer* weights is called a *minimizing* sequence of the *m*-ary tree *T* in the class *M* ($P_{\min} \in M$) if $P_{\min}$ produces the minimal Huffman cost of the *m*-ary tree *T* over all sequences from *M*, i.e.,

$$E(T,P_{\min}) \le E(T,P) \forall P \in M.$$

**Definition**. A non-decreasing sequence of positive integer weights $P = \{p_1, p_2, \ldots, p_n\}$ is called *absolutely ordered* if the intermediate sequences of weights produced by *m*-ary Huffman algorithm for initial sequence *P* satisfy the following conditions

$$p_m^{(i)} < p_{m+1}^{(i)},\ i = \overline{0, N-2}. \quad (3)$$

For an absolutely ordered sequence the equality-inequality relation (2) is transformed to the (strict) equality relation

$$p_1^{(i)} + p_2^{(i)} + \ldots + p_m^{(i)} < p_{2m}^{(i)}, i = \overline{0, N-2}. \quad (4)$$

**Lemma 1**. A *minimizing* absolutely ordered sequence of size $n = N*(m-1) + 1$ for the elongated *m*-ary tree ($m > 1$) is

$$Pmin_{\text{abs}}(N, m) = \{ Q_0(m), \underbrace{Q_1(m), \ldots, Q_1(m)}_{(m-1)\text{ times}}, \underbrace{Q_2(m), \ldots, Q_2(m)}_{(m-1)\text{ times}}, \ldots, \underbrace{Q_N(m), \ldots, Q_N(m)}_{(m-1)\text{ times}} \},$$

where $Q_0(m) = 1$, $Q_1(m) = 1$, $Q_2(m) = 2$, $Q_i(m) = Q_{i-1}(m) + (m-1)*Q_{i-2}(m)$ when $i = \overline{2, N}$.

**Proof**. Taking into account (3) (4) we obtain the following configurations of *m*-ary Huffman algorithm steps for absolutely ordered sequence of the elongated (left-sided) *m*-ary tree.

**Step 0 (Initial):**

Step 0: $p_1, p_2, \ldots, p_m, p_{m+1}, \ldots, p_{2m-1}, p_{2m}, p_{2m+1}, \ldots, p_n$; $\quad p_m < p_{m+1}$;

**Steps 1-(*N*-2):**

Step 1: $p_{m+1}, \ldots, p_{2m-1}, \sum_{j=1}^{m} p_j, p_{2m}, \ldots, p_n$; $\quad \sum_{j=1}^{m} p_j < p_{2m}$;

Step 2: $p_{2m}, \ldots, p_{3m-2}, \sum_{j=1}^{2m-1} p_j, p_{3m-1}, \ldots, p_n$; $\quad \sum_{j=1}^{2m-1} p_j < p_{3m-1}$;

Step 3: $p_{3m-1}, \ldots, p_{4m-3}, \sum_{j=1}^{3m-2} p_j, p_{4m-2}, \ldots, p_n$; $\quad \sum_{j=1}^{3m-2} p_j < p_{4m-2}$;

…

Step *i*: $p_{i*(m-1)+2}, \ldots, p_{(i+1)*(m-1)+1}, \sum_{j=1}^{i*(m-1)+1} p_j, p_{(i+1)*(m-1)+2}, \ldots, p_n$; $\quad \sum_{j=1}^{i*(m-1)+1} p_j < p_{(i+1)*(m-1)+2}$;

…

Step *N*-2: $p_{(N-2)*(m-1)+2}, \ldots, p_{(N-1)*(m-1)+1}, \sum_{j=1}^{(N-2)*(m-1)+1} p_j, p_{(N-1)*(m-1)+2}, \ldots, p_n$; $\quad \sum_{j=1}^{(N-2)*(m-1)+1} p_j < p_{(N-1)*(m-1)+2}$;

**Steps (*N*-1), *N*:**

Step *N*-1: $p_{(N-1)*(m-1)+2}, \ldots, p_{N*(m-1)+1}, \sum_{j=1}^{(N-1)*(m-1)+1} p_j$ (Note. According (1) $(N*(m-1)+1) = n$);

Step *N*: $\sum_{j=1}^{N*(m-1)+1} p_j = \sum_{j=1}^{n} p_j$.

On ***Step 0*** merging *m* leaves labeled by integers $p_1, p_2, \ldots, p_m$ are merged. Because $Pmin_{abs}(N, m)= \{p_1, p_2, \ldots, p_n\}$ is *minimizing* sequence of positive *integer* values, $p_1, p_2, \ldots, p_m$ should have minimal positive integer values, i.e., at least they must be equal. So, we can write as follows

$$p_1 = q_0,$$
$$p_2 = \ldots = p_m = q_1;$$
$$q_0 = q_1.$$

On ***Step 1*** merging (*m*-1) leaves labeled by integers $p_{m+1}, \ldots, p_{2m-1}$ and one nonleaf are merged. Again, because $Pmin_{abs}(N, m)= \{p_1, p_2, \ldots, p_n\}$ is *minimizing* sequence of positive *integer* values, $p_{m+1}, \ldots, p_{2m-1}$ should have minimal *possible* positive integer values, i.e., at least they must be equal. So, we can write as follows

$$p_{m+1} = \ldots = p_{2m-1} = q_2.$$

In the same manner for ***Steps 2,…, (N-1)*** we obtain

$$p_{2m} = \ldots = p_{3m-2} = q_3;$$
$$P_{3m-1} = \ldots = p_{4m-3} = q_4;$$
$$\ldots$$
$$p_{i*(m-1)+2} = \ldots = p_{(i+1)*(m-1)+1} = q_{i+1};$$
$$\ldots$$
$$p_{(N-1)*(m-1)+2} = \ldots = p_{N*(m-1)+1} = q_N.$$

So, the configurations of *m*-ary Huffman algorithm steps for absolutely ordered sequence of the elongated (left-sided) *m*-ary tree are transformed as follows.

**Step 0 (Initial):**

Step 0: $q_0, \underbrace{q_1,\dots,q_1}_{(m-1)\text{ times}}, \underbrace{q_2,\dots,q_2}_{(m-1)\text{ times}}, \dots, \underbrace{q_N,\dots,q_N}_{(m-1)\text{ times}}$ $\qquad q_1 < q_2$;

**Steps 1-(*N*-2):**

Step 1: $\underbrace{q_2,\dots,q_2}_{(m-1)\text{ times}}, q_0 + (m\text{-}1)\sum_{j=1}^{1} q_j, \underbrace{q_3,\dots,q_3}_{(m-1)\text{ times}}, \dots, \underbrace{q_N,\dots,q_N}_{(m-1)\text{ times}}$; $\qquad q_0 + (m\text{-}1)\sum_{j=1}^{1} q_j < q_3$;

Step 2: $\underbrace{q_3,\dots,q_3}_{(m-1)\text{ times}}, q_0 + (m\text{-}1)\sum_{j=1}^{2} q_j, \underbrace{q_4,\dots,q_4}_{(m-1)\text{ times}}, \dots, \underbrace{q_N,\dots,q_N}_{(m-1)\text{ times}}$; $\qquad q_0 + (m\text{-}1)\sum_{j=1}^{2} q_j < q_4$;

Step 3: $\underbrace{q_4,\dots,q_4}_{(m-1)\text{ times}}, q_0 + (m\text{-}1)\sum_{j=1}^{3} q_j, \underbrace{q_5,\dots,q_5}_{(m-1)\text{ times}}, \dots, \underbrace{q_N,\dots,q_N}_{(m-1)\text{ times}}$; $\qquad q_0 + (m\text{-}1)\sum_{j=1}^{3} q_j < q_5$;

…

Step *i*: $\underbrace{q_{i+1},\dots,q_{i+1}}_{(m-1)\text{ times}}, q_0 + (m\text{-}1)\sum_{j=1}^{i} q_j, \underbrace{q_{i+2},\dots,q_{i+2}}_{(m-1)\text{ times}}, \dots, \underbrace{q_N,\dots,q_N}_{(m-1)\text{ times}}$; $\qquad q_0 + (m\text{-}1)\sum_{j=1}^{i} q_j < q_{i+2}$;

…

Step *N*-2: $\underbrace{q_{N-1},\dots,q_{N-1}}_{(m-1)\text{ times}}, q_0 + (m\text{-}1)\sum_{j=1}^{N-2} q_j, \underbrace{q_N,\dots,q_N}_{(m-1)\text{ times}}$; $\qquad q_0 + (m\text{-}1)\sum_{j=1}^{N-2} q_j < q_N$;

**Steps (*N*-1), *N*:**

Step *N*-1: $\underbrace{q_N,\dots,q_N}_{(m-1)\text{ times}}, q_0 + (m\text{-}1)\sum_{j=1}^{N-1} q_j$;

Step *N*: $q_0 + (m\text{-}1)\sum_{j=1}^{N} q_j$.

Because $Pmin_{\text{abs}}(N, m)$= { $q_0, \underbrace{q_1,\dots,q_1}_{(m-1)\text{ times}}, \underbrace{q_2,\dots,q_2}_{(m-1)\text{ times}}, \dots, \underbrace{q_N,\dots,q_N}_{(m-1)\text{ times}}$ } is *minimizing* sequence of positive *integer* values, $q_0$ and $q_1$ should have minimal positive integer values, and $q_2$ should have minimal *possible* positive integer value. So, we have

$$q_0 = 1, \tag{5}$$
$$q_1 = 1, \tag{6}$$

and, taking into account (3)

$$q_2 = q_1+1 = 2, \tag{7}$$
$$q_i = (m\text{-}1)\sum_{j=1}^{i-2} q_j + 1, \text{ when } i = \overline{2, N}.$$

Consider, $q_i$ - $q_{i-1}$ when $i = \overline{3, N}$.

$$q_i - q_{i-1} = ((m\text{-}1)\sum_{j=1}^{i-2} q_j + 1) - ((m\text{-}1)\sum_{j=1}^{i-3} q_j + 1) = (m\text{-}1)\, q_{i-2}. \tag{8}$$

i.e.

$$q_i = q_{i-1} + (m\text{-}1)\, q_{i-2}, \text{ when } i = \overline{2, N}.$$

From (5), (6), (7) and (8) we obtain that $q_i$ is a function of $m$, i.e., $q_i = Q_i(m)$ and thus

$$Q_0(m) = 1,\ Q_1(m) = 1,\ Q_2(m) = 2,\ Q_i(m) = Q_{i-1}(m) + (m\text{-}1)*Q_{i-2}(m) \text{ when } i = \overline{2, N}. \tag{9}$$

The lemma has been proved.■

## 2.2. Fibonacci-like polynomials

From Lemma 1 we can see that *m*-ary Huffman tree ($m > 1$) is connected with polynomials $Q_i(m)$ (9). From that we can told that ($m$+1)-ary Huffman tree ($m > 0$) is connected with polynomials

$$G_0(m) = 1,\ G_1(m) = 1,\ G_2(m) = 2,\ G_i(m) = G_{i-1}(m) + m*G_{i-2}(m) \text{ when } i = \overline{2, N}.$$

So, we have polynomials that are defined by the recurrence relation

$$G_i(x) = G_{i-1}(x) + x*G_{i-2}(x) \text{ when } i > 2;$$

with

$$G_0(x) = 1,\ G_1(x) = 1,\ G_2(x) = 2.$$

Thus, Lemma 1 can be reformulate as

**<u>Theorem 1</u>**. A *<u>minimizing</u>* absolutely ordered sequence of size $n = N*(m-1) + 1$ for the elongated *m*-ary Huffman tree ($m > 1$) is

$$Pmin_{\text{abs}}(N, m) = \{\, G_0(m-1), \underbrace{G_1(m-1),\ldots,G_1(m-1)}_{(m-1)\text{ times}}, \underbrace{G_2(m-1),\ldots,G_2(m-1)}_{(m-1)\text{ times}},\ldots, \underbrace{G_N(m-1),\ldots,G_N(m-1)}_{(m-1)\text{ times}} \},$$

where $G_0(m) = 1$, $G_1(m) = 1$, $G_2(m) = 2$, $G_i(m) = G_{i-1}(m) + m*G_{i-2}(m)$ when $i = \overline{2, N}$.

Huffman related polynomials $G_i(x)$ are Fibonacci-like ones in contrast to Fibonacci polynomials that are defined by another recurrence relation [4]

$$F_i(x) = x*F_{i-1}(x) + F_{i-2}(x) \text{ when } x > 2;$$

with

$$F_1(x) = 1,\ F_2(x) = x.$$

The first few Fibonacci-like (Huffman related) polynomials are

$G_0(x) = 1$
$G_1(x) = 1$
$G_2(x) = 2$
$G_3(x) = x + 2$
$G_4(x) = 3x + 2$
$G_5(x) = x^2 + 5x + 2$
$G_6(x) = 4x^2 + 7x + 2$
$G_7(x) = x^3 + 9x^2 + 9x + 2$
$G_8(x) = 5x^3 + 16x^2 + 11x + 2$
$G_9(x) = x^4 + 14x^3 + 25x^2 + 13x + 2$
$G_{10}(x) = 6x^4 + 30x^3 + 36x^2 + 15x + 2$
$G_{11}(x) = x^5 + 20x^4 + 55x^3 + 49x^2 + 17x + 2$
$G_{12}(x) = 7x^5 + 50x^4 + 91x^3 + 64x^2 + 19x + 2$
$G_{13}(x) = x^6 + 27x^5 + 105x^4 + 140x^3 + 81x^2 + 21x + 2$
$G_{14}(x) = 8x^6 + 77x^5 + 196x^4 + 204x^3 + 100x^2 + 23x + 2$
$G_{15}(x) = x^7 + 35x^6 + 182x^5 + 336x^4 + 285x^3 + 121x^2 + 25x + 2$
$G_{16}(x) = 9x^7 + 112x^6 + 378x^5 + 540x^4 + 385x^3 + 144x^2 + 27x + 2$
$G_{17}(x) = x^8 + 44x^7 + 294x^6 + 714x^5 + 825x^4 + 506x^3 + 169x^2 + 29x + 2$
$G_{18}(x) = 10x^8 + 156x^7 + 672x^6 + 1254x^5 + 1210x^4 + 650x^3 + 196x^2 + 31x + 2$
$G_{19}(x) = x^9 + 54x^8 + 450x^7 + 1386x^6 + 2079x^5 + 1716x^4 + 819x^3 + 225x^2 + 33x + 2$
$G_{20}(x) = 11x^9 + 210x^8 + 1122x^7 + 2640x^6 + 3289^{x5} + 2366x^4 + 1015x^3 + 256^{x2} + 35x + 2.$

The Fibonacci-like polynomials $G_i(x)$ are normalized, i.e.

$$G_i(1) = Fib_{i+1}, \tag{10}$$

where $Fib_i$ is *i*-th Fibonacci number.

According to Theorem 1 the sequence $G_0(m), \underbrace{G_1(m),...,G_1(m)}_{(m-1)\text{ times}}, \underbrace{G_2(m),...,G_2(m)}_{(m-1)\text{ times}},..., \underbrace{G_N(m),...,G_N(m)}_{(m-1)\text{ times}}$ is minimizing absolutely ordered sequence of size $n = N*(m-1) + 1$ for the elongated ($m$+1)-ary Huffman tree ($m > 0$).

**Definition**. Sequence $G_0(m)$, $G_1(m)$, $G_2(m)$,   , $G_N(m)$ is called a *representative* Huffman $m$-sequence, i.e., representative sequence of the elongated ($m$+1)-ary Huffman tree ($m > 0$).

Several examples of representative Huffman $m$-sequences are shown in Table 1.

**Table 1**. Samples of representative Huffman $m$-sequences

| **m** | **$G_i(m)$: $i$ = 0, 1, 2,…, 12** | | | | | | | | | | | | | |
|---|---|---|---|---|---|---|---|---|---|---|---|---|---|---|
| | **0** | **1** | **2** | **3** | **4** | **5** | **6** | **7** | **8** | **9** | **10** | **11** | **12** | **13** |
| **1** | 1 | 1 | 2 | 3 | 5 | 8 | 13 | 21 | 34 | 55 | 89 | 144 | 233 | 377 |
| **2** | 1 | 1 | 2 | 4 | 8 | 16 | 32 | 64 | 128 | 256 | 512 | 1024 | 2048 | 4096 |
| **3** | 1 | 1 | 2 | 5 | 11 | 26 | 59 | 137 | 314 | 725 | 1667 | 3842 | 8843 | 20369 |
| **4** | 1 | 1 | 2 | 6 | 14 | 38 | 94 | 246 | 622 | 1606 | 4094 | 10518 | 26894 | 68966 |
| **5** | 1 | 1 | 2 | 7 | 17 | 52 | 137 | 397 | 1082 | 3067 | 8477 | 23812 | 66197 | 185257 |
| **6** | 1 | 1 | 2 | 8 | 20 | 68 | 188 | 596 | 1724 | 5300 | 15644 | 47444 | 141308 | 425972 |
| **7** | 1 | 1 | 2 | 9 | 23 | 86 | 247 | 849 | 2578 | 8521 | 26567 | 86214 | 272183 | 875681 |
| **8** | 1 | 1 | 2 | 10 | 26 | 106 | 314 | 1162 | 3674 | 12970 | 42362 | 146122 | 485018 | 1653994 |
| **9** | 1 | 1 | 2 | 11 | 29 | 128 | 389 | 1541 | 5042 | 18911 | 64289 | 234488 | 813089 | 2923481 |
| **10** | 1 | 1 | 2 | 12 | 32 | 152 | 472 | 1992 | 6712 | 26632 | 93752 | 360072 | 1297592 | 4898312 |
| **11** | 1 | 1 | 2 | 13 | 35 | 178 | 563 | 2521 | 8714 | 36445 | 132299 | 533194 | 1988483 | 7853617 |
| **12** | 1 | 1 | 2 | 14 | 38 | 206 | 662 | 3134 | 11078 | 48686 | 181622 | 765854 | 2945318 | 12135566 |
| **13** | 1 | 1 | 2 | 15 | 41 | 236 | 769 | 3837 | 13834 | 63715 | 243557 | 1071852 | 4238093 | 18172169 |
| **14** | 1 | 1 | 2 | 16 | 44 | 268 | 884 | 4636 | 17012 | 81916 | 320084 | 1466908 | 5948084 | 26484796 |
| **15** | 1 | 1 | 2 | 17 | 47 | 302 | 1007 | 5537 | 20642 | 103697 | 413327 | 1968782 | 8168687 | 37700417 |

## 2.3. Some properties of Fibonacci-like polynomials

Let

$$S(N, m) = \sum_{i=0}^{N} G_i(m) . \qquad (11)$$

Calculate $S(N, m)$. Consider

$$(m+1)* S(N, m) = (m+1)* \sum_{i=0}^{N} G_i(m)$$

$$= \sum_{i=0}^{N} G_i(m) + m* \sum_{i=0}^{N} G_i(m)$$

$$= (G_0(x) + G_1(x) + \sum_{i=2}^{N} G_i(m)) + (m*G_0(x) + m* \sum_{i=1}^{N} G_i(m))$$

$$= (m+1)*G_0(x) + G_1(x) + \sum_{i=1}^{N-1} (G_{i+1}(m) + m * G_i(m)) + m*G_N(m)$$

$$= (m+1)*G_0(x) + G_1(x) + \sum_{i=1}^{N-1} G_{i+2}(m) + m*G_N(m)$$

$$= (m+1)*G_0(x) + G_1(x) + \sum_{i=3}^{N+1} G_i(m) + m*G_N(m)$$

$$= (m+1)*G_0(x) + G_1(x) + \sum_{i=0}^{N+1} G_i(m) - (G_0(x) + G_1(x) + G_2(x)) + m*G_N(m)$$

$$= m*G_0(x) - G_2(x) + \sum_{i=0}^{N} G_i(m) + G_{N+1}(m) + m*G_N(m)$$

$$= m*1 - 2 + S(N, m) + G_{N+2}(m)$$

$$= S(N, m) + G_{N+2}(m) + m - 2.$$

So,

$$(m+1)* S(N, m) = S(N, m) + G_{N+2}(m) + m - 2, \text{ i.e.,}$$

$$m*S(N, m) = G_{N+2}(m) + m - 2.$$

Therefore

$$S(N, m) = \frac{G_{N+2}(m) + m - 2}{m} = \frac{G_{N+2}(m) - 2}{m} + 1. \qquad (12)$$

In particular,

$$S(N, 1) = \frac{G_{N+2}(1) + 1 - 2}{1} = G_{N+2}(1) - 1 = Fib_{N+3} - 1.$$

## 2.4. Cost of minimizing absolutely ordered sequence of the elongated *m*-ary Huffman tree

**Theorem 2**. The cost (i.e., weighted external path length) of elongated $m$-ary Huffman tree $T$ of size $n = N*(m-1) + 1$ for the minimizing absolutely ordered sequence $Pmin_{abs}(N, m)$ is

$$E(T, Pmin_{abs}(N, m)) = \frac{G_{N+4}(m-1) - 2}{m-1} - (N+3) = \frac{G_{\frac{n+4m-5}{m-1}}(m-1) - (n+3m-2)}{m-1}.$$

**Proof**. Let $Pmin_{abs}(N, m) = \{p_1, p_2, \ldots, p_n\}$ be the minimizing $k$-ordered sequence of the elongated binary tree $T$ of size $n$.

According to Theorem 1

$$Pmin_{abs}(N, m) = \{ G_0(m-1), \underbrace{G_1(m-1),\ldots,G_1(m-1)}_{(m-1)\text{ times}}, \underbrace{G_2(m-1),\ldots,G_2(m-1)}_{(m-1)\text{ times}}, \ldots, \underbrace{G_N(m-1),\ldots,G_N(m-1)}_{(m-1)\text{ times}} \}.$$

Weighted external path length $E(T, Pmin_{n,k})$ is

$$E(T, Pmin_{abs}(N, m)) = \sum_{i=1}^{n} l_i p_i .$$

where $l_i$ is the length of the path from the root to leaf $i$.

$T$ is the elongated binary tree, therefore

$$E(T, Pmin_{abs}(N, m)) = \sum_{i=1}^{n} l_i p_i$$

$$= N*G_0(m-1) + (m-1)\sum_{i=1}^{N} (N - i + 1) * G_i(m-1)$$

$$= N*G_0(m-1) + (m-1)\sum_{i=0}^{N} (N - i + 1) * G_i(m-1) - (m-1)*(N+1)*G_0(m-1)$$

$$= (m-1)\sum_{i=0}^{N} (N - i + 1) * G_i(m-1) - (m-2)*(N+1)*G_0(m-1) - G_0(m-1)$$

$$= (m-1)\sum_{i=0}^{N}\sum_{j=0}^{N-i} G_i(m-1) - (m-2)*(N+1)*G_0(m-1) - G_0(m-1)$$

$$= (m-1)\sum_{j=0}^{N}\sum_{i=0}^{j} G_i(m-1) - (m-2)*(N+1)*G_0(m-1) - G_0(m-1).$$

Thus, taking into account (11) and (12), we obtain

$$E(T, Pmin_{\text{abs}}(N, m)) = (m\text{-}1)\sum_{j=0}^{N}\sum_{i=0}^{j} G_i(m-1) - (m-2)*(N+1)*G_0(m-1) - G_0(m-1)$$

$$= (m-1)\sum_{i=0}^{N} S(i, m-1) - (m-2)*(N+1)*G_0(m-1) - G_0(m-1)$$

$$= (m-1)\sum_{i=0}^{N} \frac{G_{i+2}(m-1) + (m-1) - 2}{m-1} - (m-2)*(N+1)*G_0(m-1) - G_0(m-1)$$

$$= \sum_{i=0}^{N} (G_{i+2}(m-1) + (m-1) - 2) - (m-2)*(N+1)*G_0(m-1) - G_0(m-1)$$

$$= \sum_{i=0}^{N} G_{i+2}(m-1) + (m-3)*(N+1) - (m-2)*(N+1)*G_0(m-1) - G_0(m-1)$$

$$= \sum_{i=0}^{N} G_{i+2}(m-1) + (m-3)*(N+1) - (m-2)*(N+1) - 1$$

$$= \sum_{i=0}^{N} G_{i+2}(m-1) - (N+1) - 1$$

$$= \sum_{i=0}^{N} G_{i+2}(m-1) - (N+2)$$

$$= \sum_{i=0}^{N+2} G_i(m-1) - (G_0(m-1) + G_1(m-1)) - (N+2)$$

$$= S(N+2, m-1) - 2 - (N+2)$$

$$= S(N+2, m-1) - (N+4)$$

$$= \frac{G_{N+4}(m-1) - 2}{m-1} + 1 - (N+4)$$

$$= \frac{G_{N+4}(m-1) - 2}{m-1} - (N+3)$$

In particular, taking into account (10) we have for binary (i.e., $m = 2$, $n = N+1$) Huffman elongated tree

$$E(T, Pmin_{\text{abs}}(N, 2)) = \frac{G_{N+4}(1) - 2}{1} - (N+3)$$

$$= G_{N+4}(1) - 2 - (N+3)$$

$$= G_{N+4}(1) - (N+5)$$

$$= G_{N+4}(1) - (N+5)$$

$$= Fib_{N+5} - (N+5)$$

$$= Fib_{n+4} - (n+4).$$

Several examples of costs for several elongated *m*-ary Huffman trees are shown in Table 2.

**Table 2**. Costs for several elongated *m*-ary Huffman trees

| **Arity** | ***N*** (number of non-leaves in an elongated Huffman tree) | | | | | | | | | |
|---|---|---|---|---|---|---|---|---|---|---|
| ***m*** | **1** | **2** | **3** | **4** | **5** | **6** | **7** | **8** | **9** | **10** |
| 2 | 2 | 6 | 13 | 25 | 45 | 78 | 132 | 220 | 363 | 595 |
| 3 | 3 | 10 | 25 | 56 | 119 | 246 | 501 | 1012 | 2035 | 4082 |
| 4 | 4 | 14 | 39 | 97 | 233 | 546 | 1270 | 2936 | 6777 | 15619 |
| 5 | 5 | 18 | 55 | 148 | 393 | 1014 | 2619 | 6712 | 17229 | 44122 |
| 6 | 6 | 22 | 73 | 209 | 605 | 1686 | 4752 | 13228 | 37039 | 103235 |
| 7 | 7 | 26 | 93 | 280 | 875 | 2598 | 7897 | 23540 | 70983 | 212290 |
| 8 | 8 | 30 | 115 | 361 | 1209 | 3786 | 12306 | 38872 | 125085 | 397267 |
| 9 | 9 | 34 | 139 | 452 | 1613 | 5286 | 18255 | 60616 | 206737 | 691754 |
| 10 | 10 | 38 | 165 | 553 | 2093 | 7134 | 26044 | 90332 | 324819 | 1137907 |
| 11 | 11 | 42 | 193 | 664 | 2655 | 9366 | 35997 | 129748 | 489819 | 1787410 |
| 12 | 12 | 46 | 223 | 785 | 3305 | 12018 | 48462 | 180760 | 713953 | 2702435 |
| 13 | 13 | 50 | 255 | 916 | 4049 | 15126 | 63811 | 245432 | 1011285 | 3956602 |
| 14 | 14 | 54 | 289 | 1057 | 4893 | 18726 | 82440 | 325996 | 1397847 | 5635939 |
| 15 | 15 | 58 | 325 | 1208 | 5843 | 22854 | 104769 | 424852 | 1891759 | 7839842 |
| 16 | 16 | 62 | 363 | 1369 | 6905 | 27546 | 131242 | 544568 | 2513349 | 10682035 |
| 17 | 17 | 66 | 403 | 1540 | 8085 | 32838 | 162327 | 687880 | 3285273 | 14291530 |
| 18 | 18 | 70 | 445 | 1721 | 9389 | 38766 | 198516 | 857692 | 4232635 | 18813587 |
| 19 | 19 | 74 | 489 | 1912 | 10823 | 45366 | 240325 | 1057076 | 5383107 | 24410674 |
| 20 | 20 | 78 | 535 | 2113 | 12393 | 52674 | 288294 | 1289272 | 6767049 | 31263427 |
| 21 | 21 | 82 | 583 | 2324 | 14105 | 60726 | 342987 | 1557688 | 8417629 | 39571610 |